# Quantum nonlocality: no, yes, how and why.

Alejandro A. Hnilo

*CEILAP, Centro de Investigaciones en Láseres y Aplicaciones, UNIDEF (MINDEF-CONICET);
CITEDEF, J.B. de La Salle 4397, (1603) Villa Martelli, Argentina.
email: alex.hnilo@gmail.com*

March 9th, 2026.

The problem of the existence of nonlocal effects in Quantum Mechanics is discussed. The problem is divided in two: the first ("soft") one is to explain the violation of Bell's inequalities as a statistical magnitude. This can be achieved by a simple model within non-Boolean Locality and Realism. This result shows that quantum non-Locality as a consequence of the statistical violation of Bell's inequalities is inexistent. The second ("hard") problem is to explain the violation as it is calculated from series of detection outcomes. L.Sica has demonstrated that, in order to violate Bell's inequalities, the series recorded at (say) Bob when the setting at station Alice is $\alpha$, can be different from the series that would have been recorded at Bob if that setting had been $\alpha'$ instead. Therefore, non-Locality in the series of detection outcomes does exist. It cannot be experimentally verified because of its counterfactual nature, but is observed in computer simulations. An appropriate computer code is based on the simple model mentioned plus a contextual instruction. It explains "how" (Sica's) non-Locality arises, and solves the hard problem. "Why" the contextual instruction exists is explained by Hellwig and Kraus' postulate of covariant quantum state collapse. In consequence, (Sica's) non-Locality is not in contradiction with Relativity but, quite the opposite, it is implied by Relativistic covariance.

## 1. Introduction.

The discussion on the existence of quantum non-locality can be traced back to the 1935 EPR paradox and the famous "spooky action-at-a-distance". It is worth recalling that both Einstein and Bohr (the two main names in the controversy) immediately rejected the possibility of non-local effects. The issue became definite with the formulation of Bell's inequalities (BI), which made clear that Quantum Mechanics (QM) is incompatible with naïf forms of Locality and Realism. Accepting the existence of nonlocal effects implies a possible contradiction between QM and the theory of Relativity. This contradiction is usually circumvented by assuming that the limit of *c* (the speed of light) imposed by Relativity applies to propagation of information, or *signals*, and that quantum nonlocal effects cannot be used to send signals (thus the importance of "non-signaling" theorems in the field of Quantum Information). Yet, the assumption that *c* is a limit that applies only to signals is arguable, and the tension remains. A.Shimony famously stated that QM and Relativity are in "peaceful coexistence", a term which had been coined to describe the fragile relationship between the East and West blocks during the Cold War. It would be desirable ensuring that quantum non-Locality and Relativity are compatible even without the no-signaling assumption.

Nowadays, it is customary identifying violation of BI with non-Locality [1,2]. On the other hand, many distinguished researchers have insisted, with different arguments, that the belief in quantum non-Locality is misled [3-8]. A review published in 2020 cited more than 70 articles pointing out that [9]. The Copenhagen interpretation (say, the "orthodox" version of QM) gives up Realism, not Locality. Whether quantum non-Local effects exist or not is therefore an open problem. Its solution has impact not only on the foundations of Physics. If non-Local effects existed, a most sought definition of randomness (and a reliable source of random numbers) would be at hand [10].

This paper discusses in which sense quantum non-Locality does not exist, in which sense it does exist, how it is visible in a numerical simulation, and why it is compatible with Relativity, independently of the no-signaling assumption.

The discussion involves the setup sketched in Figure 1. Source S emits pairs of photons entangled in polarization, which are detected at distant stations A and B after polarizers oriented at angles $\{\alpha,\beta\}$. Detections' time values are stored (time-stamping). F.ex., detector "-1" in A and "1" in B fired during time slot #101. Note that number of single detections is equal to number of coincidences, i.e., ideal efficiency is supposed here. No detector fired at time slots #100 and #103. In an actual experiment, the number of slots with "zero" is much larger than the others.

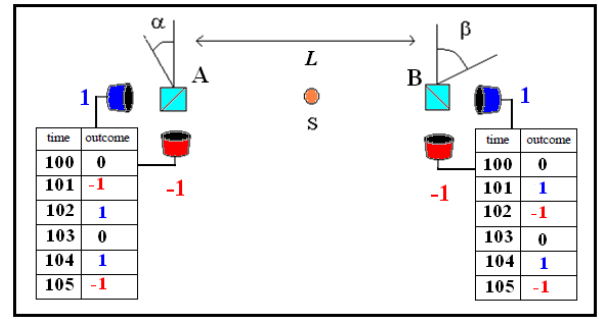

Figure 1: Sketch of an ideal time-stamped Bell's experiment, efficiency is equal to 1.

In the next Section, "Bell's non-Locality" (that is, non-Locality based on the violation of BI, which involves *statistical* magnitudes) is shown to be inexistent. The following Section deals with a different form of non-Locality, which involves individual detections, or *series* of outcomes (as the ones recorded in Fig.1). It is named here "Sica's non-Locality" and it does exist, but involves a counterfactual situation. The idea of two different types of quantum non-Locality, namely, one related with "subquantum statistical independence" based on the (statistical) violation of BI, and other based on (individual detections) Lüder's postulate, has been noted by A.Khrennikov [11]. Section 4 reviews a computer code that not only reproduces the statistical QM predictions successfully, (the "soft" problem) but that also allows predicting when and where a photon is detected (the "hard" problem). It allows exploring counterfactual situations and numerically confirms Sica's non-Locality. This code includes a contextual instruction. Its existence is explained in Section 5, as a consequence of Hellwig and Kraus' postulate of quantum state's *covariant collapse* [12]. This feature ensures there is no contradiction between (Sica's) quantum non-Locality and Relativity, with no need of the non-signaling assumption.

## 2. Bell's non-locality does not exist.

The title of this section is demonstrated with a counter-example. A crucial step in the derivation of BI is the statistical independence of probability detections at the hidden variables $\lambda$ (or "subquantum" [11]) level:

$$P_{AB}{}^{ab}(\alpha,\beta,\lambda) = P_A{}^a(\alpha,\lambda).P_B{}^b(\beta,\lambda) \qquad (1)$$

where A,B refers to the stations in Fig.1, and *a*, *b* are the possible outcomes: $a = \pm 1$, $b = \pm 1$ [13]. Statistical independence is implied by Locality. In order to make impossible any interaction between A and B at subluminal speed, major experimental efforts have been performed. They have required distant stations with arbitrary and fast-changing values of the settings $\{\alpha,\beta\}$ [14-19]. The BI were observed to be violated even in these conditions, so that many researchers concluded the equality in Eq.1 to be false, and hence, that non-Local effects exist.

Nevertheless, classical probabilities as in eq.1 presuppose a Boolean algebra [20-22]. In fact, BI are equivalent to Boole's conditions of completeness. Violation of BI with Boolean variables is hence a logical impossibility. If non-Boolean hidden variables are assumed instead, the violation of BI is trivially explained. This explanation is discussed at length in [23]; the essential steps are reviewed next.

The simplest non-Boolean description of the experiment in Fig.1 uses vectors in real space as the hidden variables. Systems having more than one feature are not defined as the intersection of the corresponding sets (as in familiar Boolean logic), but as the projection of the corresponding vectors. In what follows, vectors are indicated with bold typing. The operation "projection" of vector **b** into vector **a** is:

$$\mathbf{a}.\mathbf{b} \equiv |b.cos(\gamma)|.\mathbf{e}_a \qquad (2)$$

where $\gamma$ is the angle between **a** and **b**, b is the modulus of **b**, and $\mathbf{e}_a$ is the unit vector in the direction of **a**. The resulting vector represents the systems that have features **b** and (then) **a**. Exclusive features, which are represented by disjoint sets in Boolean logic, are now represented by orthogonal vectors. The operation "projection" is neither commutative nor associative. It implies a non-Boolean algebra, which has paradoxical features [23,24]. In particular, the $cos(\gamma)$ factor in eq.2 anticipates the violation of BI.

The problem now is how to link vectors' modulus, which is a continuous variable, to discontinuous particle detection. A.Khrennikov has named this "the true quantum problem". Born's rule of QM is the simplest solution: the vector state's squared modulus gives the *probability* of detecting a particle. But, this solution limits QM to be a statistical description; it gives up any hope of predicting the single outcomes in Fig.1. Besides, it forces an entangled pair to be described by a single entity. Say, the state vector $|\varphi^+{}_{AB}\rangle$ = $(1/\sqrt{2}).\{|x_A\rangle.|x_B\rangle + (|y_A\rangle.|y_B\rangle\}$ in the usual notation. This entity does not exist in real space, but in an abstract 4-dimensional product space. The state vector $|\varphi^+{}_{AB}\rangle$ has no internal parts, it is an "atom" (in the original ancient Greek meaning, i.e., an entity with no internal parts) of arbitrary size. Unsurprisingly, this strange entity creates apparent "nonlocal" effects.

The next simplest solution to relate vector modulus with particle detection is to define a threshold condition [25]. Suppose then that each photon in Fig.1 carries a hidden variable (HV) named $\mathbf{V}(t)$, which is a vector in real space orthogonal to the direction of propagation. The modulus of $\mathbf{V}(t)$ is V(t); its unit vector is $\mathbf{v}(t)$, which is at an angle $\nu(t)$ with the *x*-axis. Let also assume that in a (long) experimental run of duration $Tr$, the number $N \gg 1$ of detected photons is:

$$N \approx (1/u).\int_0^{Tr} dt.|\mathbf{V}(t)|^2 \gg 1 \qquad (3)$$

where *u* is a threshold value. The vector transmitted through a polarization analyzer oriented at an angle $\alpha$ are represented by the projection of $\mathbf{V}(t)$ into the direction $\mathbf{e}_\alpha$. The detected number of particles $N_\alpha \gg 1$ after the analyzer is then:

$$N_\alpha \approx (1/u).\int_0^{Tr} dt.|\mathbf{e}_\alpha.\mathbf{v}(t).V(t)|^2 =$$

$$= (1/u).\int_0^{Tr} dt.V^2(t).cos^2[\nu(t)-\alpha] \qquad (4)$$

In a non-polarized beam there is no correlation between $\nu(t)$ and V(t), nor with $\alpha$ (which is chosen by the observer), so the time average value of the cosinus-squared factor is ½, and $N_\alpha \approx \frac{1}{2}.N$ for all values of $\alpha$, as expected.

If the photons are entangled as $|\varphi^+{}_{AB}\rangle$, then their vector-HVs are related, at the moment of their emission at the source, as $\mathbf{V}^B(t)=\mathbf{V}^A(t)$ [23]. Detections in station A are produced by the projection of $\mathbf{V}^A(t)$ into the axis $\mathbf{e}_\alpha$, i.e. $\mathbf{e}_\alpha.\mathbf{V}^A(t)$. This vector exists in a counterfactual definite way (≈ Realism). Hence, in the space of station B a vector component $e_\alpha.V^B(t)$ also exists, even if no observation is performed anywhere. The vector at station B involved in the calculation of the number of (+1,+1) coincidences $N^{++}$ is the component of $e_\alpha.V^B(t)$ that corresponds to transmission through the B-polarizer. This feature, which implies further "filtering", is given by projection into $e_\beta$. Therefore, by mere (non-Boolean) filtering, the vector component that produces coincidences is $e_\beta.e_\alpha.V^B(t)$. Then:

$$N^{++} \approx (1/u). \int_0^{Tr} dt. |e_\beta.e_\alpha.v^B(t).V^B(t)|^2 =$$

$$= (1/u). \int_0^{Tr} dt. |e_\beta.e_\alpha.cos^2[v^B(t)-\alpha].V^B(t)|^2 =$$

$$= cos^2(\alpha-\beta).(1/u.). \int_0^{Tr} dt. cos^2[v^B(t)-\alpha]. V(t)^2 =$$

$$= cos^2(\alpha-\beta).½.N \qquad (5)$$

which is the QM prediction for the state $|\varphi^+_{AB}\rangle$, and violates BI. This description (vectors in real space plus a threshold condition) can be extended to all Bell, Eberhard and GHZ states, and explains Hong-Ou-Mandel and teleportation effects [23,24]. It is not aimed to replace the QM description, but to show that an explanation of these phenomena within (non-Boolean) Locality and Realism is possible.

If Born's rule is applied to the integrands in eqs.4 and 5 with scaling $\langle V^2\rangle.Tr/u.N=1$, then the QM predicted *probabilities* are obtained. This means that vector-HVs and wavefunctions are somehow equivalent, at least in this case and from the point of view of calculating probabilities (see also Section 5).

Note that $V^B(t)$ and $V^A(t)$ are definite at the source and evolve and produce detections separately, i.e., locally. As the violation of BI is explained by a local (non-Boolean) model, it is concluded that "Bell's non-Locality" (i.e., non-Locality as a consequence of the *statistical* violation of BI) does not exist.

## 3. Sica's non-locality does exist.

The vector-HVs model solves the "soft" problem, that is, how to explain the statistical ($N \gg 1$) violation of BI (QM is another approach that does it). The "hard" problem is to explain the violation of BI from the series of outcomes in Fig.1. It involves the prediction of single detections and, at this point, it is unsolved. To study this problem, the approach due to L.Sica [26] is pertinent.

Sica's approach considers the series of binary outcomes produced in the setup of Fig.1. These series underlie any measurement in a Bell's setup. The approach is based only on arithmetical features, and applies to series of any length. It is reviewed next.

Consider the series $a_i$ and $b_i$ of ±1 outcomes, as in Fig.1. They are bits in an electronic memory or signs printed in paper. They are unquestionably *real*. For these series, the Clauser-Horne-Shimony and Holt parameter $S_{CHSH}$ [13] is:

$$N.S_{CHSH} = |\Sigma\ a_i.b_i - \Sigma\ a_i.b'_i| + |\Sigma\ a'_i.b_i + \Sigma\ a'_i.b'_i| \qquad (6)$$

where the sums are over the series' length $N$ and the primed values, say $(a'_i,b'_i)$, mean series recorded with angle settings $\{\alpha',\beta'\}$. The Clauser-Horne (CH) parameter $J_{CH}$ [13] involves only one detector per station, say the "+1" ones in Fig.1:

$$J_{CH} = \Sigma\ (a_i.b_i + a_i.b'_i + a'_i.b_i - a'_i.b'_i - a_i - b_i) = \Sigma\ T_i \qquad (7)$$

Demonstrating that $S_{CHSH} \leq 2$ and $J_{CH} \leq 0$ is simple [26,34]. Let see the first term in eq.6:

$$|\Sigma\ a_i.b_i - \Sigma\ a_i.b'_i| = |\Sigma\ a_i.(b_i - b'_i)| \leq \Sigma\ |a_i|.|b_i - b'_i| = \Sigma\ |b_i - b'_i| \qquad (8)$$

because $a_i = \pm 1\ \forall i$. In the same way, the second term in eq.6 is bounded by:

$$|\Sigma\ a'_i.b_i + \Sigma\ a'_i.b'_i| \leq \Sigma\ |b_i + b'_i| \qquad (9)$$

summing up eqs.8 and 9:

$$|\Sigma\ a_i.b_i - \Sigma\ a_i.b'_i| + |\Sigma\ a'_i.b_i + \Sigma\ a'_i.b'_i| \leq \Sigma\ |b_i - b'_i| + \Sigma\ |b_i + b'_i| \qquad (10)$$

For a given value of $i$, the first term in the rhs is 2 (0) if $b_i$ and $b'_i$ have the same (different) sign. The opposite occurs for the second term in the rhs, so that the rhs is 2 for all values of $i$. Hence:

$$(1/N).\{\ |\Sigma\ a_i.b_i - \Sigma\ a_i.b'_i| + |\Sigma\ a'_i.b_i + \Sigma\ a'_i.b'_i|\ \} \leq 2 \qquad (11)$$

That is, $S_{CHSH} \leq 2$.

In eq.7, any arbitrary term is $T_i = a_i.(b_i + b'_i) + a'_i.(b_i - b'_i) - a_i - b_i$. If $a_i = 0$ and $a'_i = 1$, then $T_i = -b'_i \leq 0$ (recall $b'_i = 0$ or 1 now). If $a_i = a'_i = 0$, then $T_i = -b_i \leq 0$. If $a_i = 1$ and $a'_i = 0$, then $T_i = b'_i - 1 \leq 0$. Finally, if $a'_i = a_i = 1$ then $T_i = b_i - 1 \leq 0$. Therefore, $T_i \leq 0\ \forall i \Rightarrow J_{CH} \leq 0$, and $J_{CH} \leq 0$ is demonstrated.

At this point a natural question arises: if the BI are demonstrated to be valid for *any* possible series of outcomes of any length, how is it possible that they are observed to be violated?

The answer is that the observation of BI requires recording series for different angle settings (say, A=$\alpha$ and A=$\alpha'$). This cannot be done during the *same* time slot. A typical assignment of settings during a run of total duration T is shown in the upper part of Figure 2. A possible table of recorded series of outcomes is shown in the lower part ($N$=4). If the settings are varied fast and randomly notation is more complicated but, as far as no (conspiratorial) communication between the stations exists, the conclusions are the same.

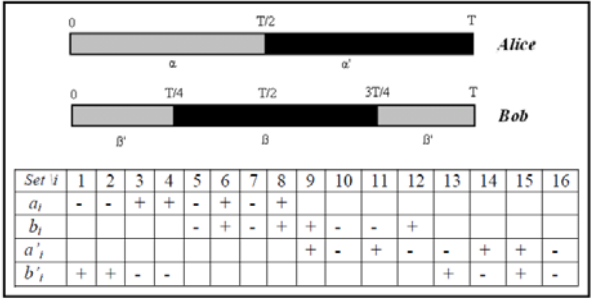

Figure 2, *Up*: Typical distribution of total recording time T among the different angle settings necessary to test the BI. F.ex: A= α between t=0 and t=T/2 (gray area in line "Alice"), B= β between t=T/4 and t=3T/4 (black area in line "Bob"). *Down*: possible series of outcomes, $S_{CHSH}$ = 2. The empty boxes correspond to non-performed observations, these outcomes are *not* "zero" (compare with Fig.1).

The empty boxes in Fig.2 correspond to non-performed observations. F.ex: for time slots from $i$=9 to 16, the setting is A=α', so that there are *no outcomes* (neither +,-, nor 0) for the elements $a_9$ to $a_{16}$, which are defined for A=α. For this reason, strictly speaking, BI cannot be calculated from the data in Fig.2. Calculation can be done only after a "possible world" is defined [27]. This may sound mysterious, but it just means that information external to the experiment must be provided. This limitation has been long ago recognized to be tacitly present, in addition to Locality and Realism, when measured data are inserted into the BI. It has received different names: ergodicity [28], homogeneous dynamics, uniform complexity, counter-factual stability, etc [29-33].Tables as the one in Fig.2 make it visually evident.

As said, the series of outcomes are unquestionably *real*. The assumption equivalent to Locality in this approach is (*Sica's Locality*):

- The series $a_i$, $a'_i$ ($b_i$, $b'_i$) are the same regardless B = β or β' (A = α or α').

In other words, the series recorded in one station do not change if the setting in the *other* station is changed. Sica's Locality defines a "possible world". In this possible world the usual BI are retrieved, as it is shown next.

Actual long series are (almost) never identical, so that the assumption above must be relaxed. F.ex, in Fig.2, $a_i$ is (-,-,+,+) (slots $i$=1 to 4, when B=β'), but it is (-,+,-,+) (slots $i$=5 to 8, when B=β). Nevertheless, if Sica's Locality is valid, then the series recorded at different times are essentially the same (as a shuffled set of playing cards is the same as it was when still packed). Therefore, it must be possible reordering the series to make them truly identical. "Reordering" here means that the sub-indexes $i$ of the series (f.ex.) $a_i$ recorded between $T/4$ and $T/2$ (i.e, when B=β) are changed so that the series becomes equal, element by element, to the $a_i$ recorded between 0 and $T/4$ (i.e, when B=β'). For long and nearly balanced series this reordering seems always possible. Yet, there is a *restriction*: the value of the correlations between outcomes in both stations cannot change, for they are physically real results. If an element in one station (say, $a_k$) changes position, the element simultaneously recorded in the other station (say, $b_k$) must change position correspondingly, so that the product $a_kb_k$ for the new value of $k$ is the same as for the old value of $k$. Because of this restriction, reordering is not always possible (see later).

When legitimate (i.e., obeying the restriction) reordering is possible the table is redundant, for all series appear twice. The same information can be then stored in a smaller, *condensed* table. F.ex: let reorder the series in Fig.2 (for which $S_{CHSH}$=2) so that $a_i$ when B=β' becomes equal to $a_i$ when B=β: (-,+,-,+). Also, so that $b_i$ when A=α' becomes equal to $b_i$ when A=α: (-,+,-,+), etc. The data can be then displayed in the table of Figure 3. Note the lines' lengths halved (from 8 to 4) and the empty boxes disappeared.

| Set\i | 1 | 2 | 3 | 4 |
|---|---|---|---|---|
| $a_i$ | - | + | - | + |
| $b_i$ | - | + | - | + |
| $a'_i$ | - | + | + | - |
| $b'_i$ | + | - | + | - |

Figure 3: The Table in Fig.2, condensed.

As demonstrated before, in a table without empty boxes the BI are valid regardless the values of the outcomes and the length of the series. Therefore: *Sica's Locality valid ⇒ series can be reordered ⇒ table can be condensed ⇒ BI are valid.* By inversion, if BI are violated (when calculated in the usual way) then the series *cannot* be reordered and Sica's Locality is *not* valid [34].

Logic suffices to demonstrate: *violation of BI ⇒ Sica's non-Locality* but, in order to understand how it works, let consider the simplest case *N*=2, see Figure 4.

| Set\ i | 1 | 2 | 3 | 4 | 5 | 6 | 7 | 8 |
|---|---|---|---|---|---|---|---|---|
| $a_i$ | - | + | - | + | | | | |
| $b_i$ | | | - | + | - | + | | |
| $a'_i$ | | | | | - | + | - | + |
| $b'_i$ | + | - | | | | | - | + |

Figure 4: Possible table of factual series (distribution of time as in the upper part of Fig.2) with $S_{CHSH}$=4, *N*=2. Sica's Locality does not hold: the series $b'_i$ at $i$=1,2 (+,-) and $i$=7,8 (-,+) are different, despite B=β' in both cases (but A is different). Legitimately reordering the series to hold to Sica's Locality is impossible.

Let suppose that $a_i$= (-,+) is measured for time slots $i$=1,2 (i.e., when B=β' and A=α), and that $b'_i$= (+,-) is simultaneously measured, so that E(α,β')= -1. In order to hold to Sica's condition, $a_i$ for $i$= 3,4 (i.e., when B= β and A=α) must be also (-,+), as shown in the Table. This does not necessarily occur in reality, but it can be achieved by reordering the elements of the measured series $a_i$. Let suppose then that in $i= 3,4$ the series $b_i$ is simultaneously measured to be (-,+), so that $E(\alpha,\beta)= 1$. In order to hold to Sica's condition, $b_i$ measured in $i= 5,6$ must be reordered (if necessary) to be (-,+) too. Let suppose that $a'_i$ is simultaneously measured to be (-,+) in $i= 5,6$ so that $E(\alpha',\beta)= 1$. At $i= 7,8$, $a'_i$ is reordered (if necessary) to (-,+) to fit Sica's condition. At this point, if $b'_i$ is simultaneously measured to be (-,+), then $E(\alpha',\beta')= 1$ and $S_{CHSH}= 4$. But, in this case, it would be $b'_i= (+,-)$ for time slots $i= 1,2$ (A= $\alpha$) and $b'_i= (-,+)$ for time slots $i= 7,8$ (A=$\alpha'$), violating Sica's condition. There is no possibility of further reordering.

Consequently, not all the series in Fig.4 ($S_{CHSH}=4$) can be made to hold to Sica's condition by legitimate reordering. Nevertheless, the difference between the series of outcomes $b'_i$ when $i=1,2$ (A=$\alpha$) and when $i=7,8$ (A=$\alpha'$) does not need a "spooky action at a distance" to be intuitively acceptable. It is understood simply because the series are measured at different times. There is no compelling reason for Sica's condition to hold for series recorded at different times. On the other hand, Sica's non-Locality is evident in the possible difference implied between an observed series (say, in station B when B=$\beta'$ and A=$\alpha$, that is, during the period from t=0 to t=T/4), and the series that would have been observed in B (also from t=0 to t=T/4 and with B=$\beta'$) if it had been A=$\alpha'$ instead. But the second series is a counterfactual, and hence unobservable. For this reason, it is clear that Sica's non-Locality cannot be used to send signals.

In Sica's approach, violation of BI is explained by giving up the assumption that legitimate reordering is always possible. In my opinion, this choice is more acceptable to intuition than giving up Locality or Realism. It just means that series recorded at different times can be essentially different from each other. In other words, it means that the "possible world" defined by Sica's Locality is not *our* world.

Note that Sica's approach deals with series of outcomes of arbitrary length, which reality is undeniable, plus simple arithmetic properties. No elaborated (and hence, vulnerable) arguments are involved. The reliability of the derived conclusions is, in consequence, hard to object.

**4. How: the WQM computer code.**

D.Graft claimed that the only way to really know "how Nature does it" (that is, *how* BI are violated) [2] is to draw a computer code that outputs series of outcomes (as the lists in Fig.1) which statistical averages violate the BI. A possible version of that code, named "WQM" was proposed in [35] (it includes a commented text of the code ready to be run in QBASIC). The essential features and results are reviewed next.

WQM follows the approach in Section 2. Each photon carries a real vector $\mathbf{V}(t)$ (this is chosen for simplicity, it might be complex as well), which is the same for both photons in the entangled pair but changes arbitrarily from one pair to the next. It is projected on the polarizers' axes at the stations. The modules of these projections are summed up in "memories" in each detector. When the sum in one memory becomes higher than a threshold value $u$, a particle is detected in the corresponding detector, and $u$ is subtracted from that memory. Therefore, although $\mathbf{V}(t)$ varies arbitrarily, the mechanism of particle detection is deterministic. If one knows $\mathbf{V}(t)$, the polarizer's orientation and the values in the memories, then one knows with certainty which detector fires and when.

A code running in a classical computer necessarily obeys a Boolean algebra. Vorob'yev theorem states that, in order to reproduce the results of a non-Boolean algebra with Boolean tools, the *context* must be defined [36]. In consequence, WQM must include a contextual instruction to fit QM predictions. The instruction is this: if a particle is detected at one of the polarizer's gates in station A, then $\mathbf{V_B}(t)$ becomes instantaneously (see Section 5) parallel to the axis of that polarizer's gate. Leaving aside this instruction, the detection mechanism at B is identical to the one at A. WQM reproduces accurately all QM predictions (see Figure 5). Besides, it solves the "hard" problem, for it predicts when and where a photon is detected.

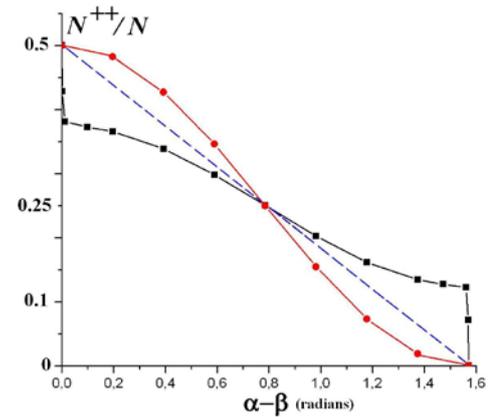

Figure 5 (from [35]): Rate of (+1,+1) coincidences as a function of the difference angle (in radians, for state $|\varphi^+_{AB}\rangle$) calculated ($N=10^6$) according to: WQM (red circles, this is also the QM prediction); the same, without the contextual instruction (black squares). The latter coincides with a semi-classical theory of radiation excepting at $\alpha-\beta= 0, \pi/2$, where it coincides with QM. The dashed blue straight line is the correlation limit imposed by the BI.

If the contextual instruction is deleted, then the curve of a semi-classical theory of radiation is obtained, excepting at the points of perfect correlation and anti-correlation, where the QM values are reproduced, see Fig.5.

Note that detections in A ("Master") influence detections in B ("Slave") through the contextual instruction, but not the other way. Nevertheless, stations A and B can be made to switch roles arbitrarily during a numerical run and WQM still reproduces all QM predictions, even for ideal efficiency and even if the angle settings are also varied arbitrarily during the run.

WQM allows exploring counterfactual situations, which are inaccessible in real experiments. Using the *same* series of values of the vector-HVs and initial values in the memories, photons are detected in the +1 or -1 gates at different time values depending of the analyzers' orientations. But, not only depending of the analyzer's orientation in the same station (this is obvious), but also depending of the analyzer's orientation in the *other* station.

F.ex., for time values between $i$=20 and $i$=40 with settings α=0, β=π/8, detections occur in station B at time values (in a specific run, see [35]):
+1 gate: 21,23,26,28,32,34,36,38,40.
-1 gate: 20,22,24,25,27,29,30,31,33,35,37,39.
If α=π/4 instead (setting β=π/8 remain the same):
+1 gate: 21,22,24,25,28,29,30,34,36,38,39,40.
-1 gate: 20,23,26,27,31,32,33,35,37.
That is, the series of outcomes in B has changed, although it is the setting *in station A* what has changed (in B, it is always β=π/8). This means that Sica's Locality is violated, as expected. The change is, of course, a consequence of the contextual instruction.

It is usual defining the HVs as intrinsically inaccessible, but let go one step further: **V**(t) may be interpreted as a representation of tiny electric fields emitted by the atoms at the source. The memories may be interpreted as a representation of collective states of the atoms in the detectors' sensitive surfaces. Knowing the values of these magnitudes is as inaccessible to an actual observer as the positions and momenta of all particles in a mol of ideal gas. Therefore, although the detection mechanism in WQM is deterministic, the outcomes in Fig.1 are unpredictable in practice.

## 5. Why: Hellwig and Kraus' postulate of covariant collapse of the quantum state.

Even if faster-than-light signaling is guaranteed to be impossible, the very idea of the existence of non-Local effects (in this case, the contextual instruction in the WQM code) seems at odds with fundamental ideas of Relativity. Conflict between QM and Relativity arises even without the BI. Consider the basic phenomenon of quantum state collapse in the space-time diagram usual in Relativity, see Figure 6. In these diagrams, an observable event occurs at *x*=0, t=0. The lower cone between blue straight lines is the absolute past of the event; the upper cone is its absolute future. The remaining regions ("elsewhere") do not belong to the event's past neither to its future. They are causally decoupled from the event. Suppose a quantum state that, at t<0, is in an eigenstate of momentum (Δ*p*=0). Then it is completely delocalized in space. It can be found anywhere inside the gray region in Fig.6 (left). If at t=0 position is measured to be *x*=0 (Δ*x*=0) then, according to standard (non-relativistic) QM, the probability of finding the state vanishes everywhere *instantaneously*, even in the space-like separated regions, which are supposed to be causally decoupled from the event. Conversely, if a state localized at *x*=0 at t<0 is measured in an eigenstate of momentum at t=0 (Fig.6, right), then the probability of finding the state becomes instantaneously different from zero everywhere, even to the point Q which is outside the future lightcone, suggesting some kind of super-luminal propagation.

These problems are known from long ago, and are consequences of the idea of instantaneous collapse. At the early stages of relativistic QM, Landau concluded that momentum at a given time and indeed all "nonlocal" properties of the quantum state at a given time cannot be valid observables for relativistic systems [37]. In 1970, Hellwig and Kraus (HK) [12] postulated that the collapse does not occur instantaneously, but along the past lightcone of the measurement event. This process is clearly covariant, since the lightcone transforms into itself under Lorenz transformations. Notably, it implies that the collapse starts a time ≈*L/2c* (where *L* is the spatial spread of the state within the cone) before the measurement event takes place. Experiments to test this *pre-collapse* effect have been proposed [38-40], but are difficult to do.

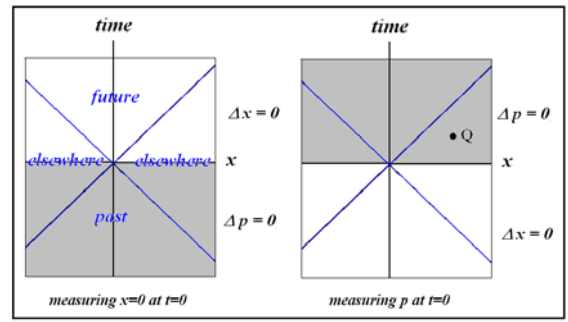

Figure 6: The problem of quantum state collapse. The gray regions are those in which the amplitude of the quantum state is different from zero if the collapse is instantaneous.

In 1981, Aharonov and Albert (AA) [41] accepted the HK postulate to describe the measurement of local observables, but found it wanting to describe "non-local" ones. AA claimed that no covariant history of the quantum state can be defined; only covariant histories of observable probabilities can be defined. The AA paper was influential, and the HK postulate was mostly forgotten. Almost 20 years later, Mould [42] refuted AA objections to the HK postulate, as well as other objections that had been raised by Cohen and Hiley [43], and d'Espagnat [44]. Elucidating the precise way or amount AA or HK are right or wrong is beyond the scope of this paper. It seems pertinent mentioning that while AA paper speaks of *particles*, the HK one speaks of *fields*.

An important feature of the HK postulate is that the collapse occurs not only in the future light cone of the measurement event, but also in the side cones ("elsewhere"). Figure 7 is the space-time diagram of a typical Bell's experiment in the reference frame where the whole setup is stationary. Source S emits one pair of photons at time *To*. The photons propagate through optical fibers (thick curves) and are detected in stations A and B at time *Td* (assumed simultaneous in this frame for simplicity). The events at A and B are space-like separated, but according to the HK postulate the collapse at A changes the quantum field in the whole shaded region in Fig.7, which includes event B. This explains *why* the contextual instruction exists without contradiction with Relativity. The contextual instruction is only apparently "instantaneous".

In 1970 the BI were not well-known yet. This is probably the reason why HK did not mention the impact of their postulate on the violation of the BI. However, they did mention that "*correlations of this type are responsible for the famous EPR paradox*" (p. 570).

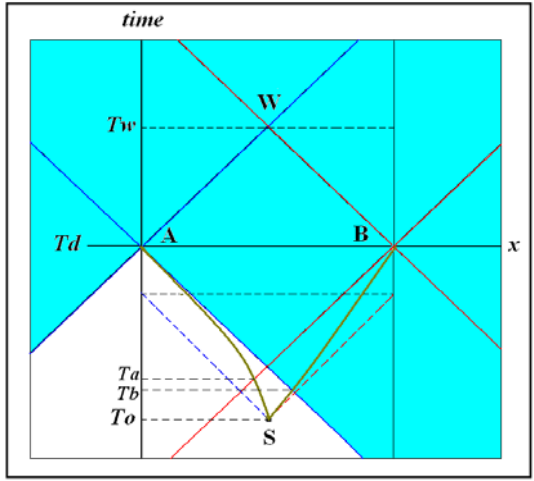

Figure 7: Space-time diagram of a typical Bell's experiment in the reference frame in which S, and A and B are stationary. The shaded areas are those in which the quantum field observed in A collapses according to the HK postulate. They include the event in B, and justify the contextual instruction in the WQM code. Symmetrically, A is into the collapsed region of B (not shown).

Note that the photon propagating towards B enters the collapsed zone of event A at time $Tb$. The photon propagating towards A enters the collapsed zone of event B at time $Tb$. In general, $Tb \neq Ta$. This difference explains the different roles (Master or Slave) played by each station in the WQM code. In Fig.7, the B-photon "knows" the result in A before the A-photon "knows" the result in B. In this case, it is natural to assume A plays Master, and B, Slave. Recall that, anyway, roles can be switched arbitrarily during a numerical run and the code still reproduces QM predictions. If the photons propagate through free space the past lightcones of both A and B intersect at S, so the information on the other side's outcome is available at emission [45]. This feature relates the HK postulate with the absorber theory of radiation by Wheeler and Feynman [46], which explains the violation of BI in an elegant way [47].

The HK postulate has the consequences shown in Fig.7 if quantum states A and B are separated and collapse separately. Separate vector-HVs are in fact assumed in the model underlying the WQM code, so that the whole picture is consistent. Assuming the HK postulate valid, WQM solves the "hard" problem and Sica's non-Locality is explained with no risk of conflict with Relativity.

Yet, there is a weak point in this picture: the HK postulate applies to quantum fields, while the **V**(t) are classical fields. Nonetheless, recall (Section 2) that correct probabilities of detection and coincidence in each gate of the polarizer can be obtained by squaring the modulus of the involved component of **V**(t). That is, by applying Born's rule as if the vector-HVs were quantum states. Therefore, operating with the **V**(t) as if they were quantum states is reasonable, at least in this case and for the purposes of calculus (WQM is just a numerical simulation). Anyway, it would be certainly desirable developing a relationship between vector-HVs and the HK postulate in a rigorous way.

On the other hand, the standard QM description assumes the quantum state to be the single, inseparable (large size "atom") $|\varphi^+_{AB}\rangle$. In this case the HK postulate implies the collapse to not to occur at A and B, but at the event W (see Fig.7), a time $Tw$ later than the photons' arrival to the stations. A similar effect has been discussed by A.Kent as one of the forms (additional to "essential" and "extended") of the *collapse-locality loophole* [48] and it is, in principle, testable.

**Summary and conclusions.**

The existence of "quantum non-locality" is discussed, and a computer code explaining "how Nature does it" is proposed.

A simple model using classical vector-HVs (which are counterfactually definite and related only at the moment of emission) in real space plus a threshold condition is reviewed. It explains the statistical violation of BI (the "soft" problem) within (non-Boolean) Local Realism. This counter-example shows that identifying violation of BI with non-Locality is misled, as it had been claimed by many distinguished researchers.

The photon-by-photon explanation of the violation of BI (the "hard" problem) is a different matter. The QM approach, because of the statistical nature of Born's rule, gives up from start any hope to solve it. An argument due to L.Sica shows that BI are violated only if the series recorded in one station (say, A) can be different from the one that would have been recorded (in A) if the setting in B had been different (say, β instead of β'). This argument implies some kind of (counterfactual, unobservable) non-Locality. Yet, its observable consequences are intuitively acceptable.

The "hard" problem is solved by the WQM code, which adds a contextual instruction to the simple model based on the vector-HVs. This code can be run with the same values of the vector-HVs and memories but different settings, allowing the exploration of counterfactual situations. It provides numerical verification of Sica's non-Locality.

In turn, the existence of the contextual instruction is explained by the HK postulate of covariant quantum collapse. This postulate guarantees there is no risk of conflict with Relativity, independently of the "no-signaling" arguments. It implies a pre-collapse effect, which test is an appealing proposal for future experiments.

In short: Bell's non-Locality does not exist. Sica's non-Locality does exist, but its observable consequences can hardly be named non-Locality. The feature that is clearly non-local involves a counterfactual and is thus experimentally unobservable. It is numerically observable by running a computer code (WQM) that includes a contextual instruction. This instruction is a consequence of the HK postulate. The complete picture shows not only that Sica's non-Locality is compatible with Relativity, but that it is even *implied* by relativistic covariance. This close relationship between quantum non-Locality and Relativity (as opposed to the fragile "peaceful coexistence" mentioned before) has been already noted by A.Suarez [49].

The approach discussed here provides a consistent description of "how Nature does it", but is not intended to replace the usual QM description. It is intended just to elucidate the meaning of quantum non-Locality, and to show its compatibility with Relativity without using the no-signaling assumption. It is also an example that "atoms" of arbitrarily large size as $|\varphi^+_{AB}\rangle$ and Born's rule are convenient calculation tools, but not inescapable features of Nature.

Of course, alternative descriptions exist. The HK postulate can be rejected or, in general, the *ontic* interpretation of the quantum state can be rejected in favor of the *epistemic* one [3,50-52]. However, in my opinion, these alternatives leave unexplained how and why (Sica's) quantum non-Locality arises, and how to solve the "hard" problem without opening a potential conflict with Relativity.

**Acknowledgments.**

This work received support from grant PIP 00484-22 from CONICET (Argentina).

**References.**